\documentclass[aps, prb, twocolumn, reprint, superscriptaddress, floatfix, citeautoscript, longbibliography]{revtex4-1}
\usepackage[T1]{fontenc} 
\usepackage[utf8]{inputenc}
\usepackage{newtxtext}
\usepackage{newtxmath}
\usepackage{graphicx}
\usepackage{siunitx}
\usepackage{booktabs}
\usepackage[english]{babel}
\usepackage{csquotes}
\usepackage{microtype}
\usepackage{chemformula}
\usepackage{multirow}
\usepackage[unicode,colorlinks=true,allcolors=blue]{hyperref}
\usepackage{mathrsfs}

\usepackage{mathtools, nccmath}

\usepackage{cleveref}

\newcommand{\bnmr}{$\beta$-NMR}

\newcommand{\eli}{\ch{$^8$Li}}
\newcommand{\elip}{\ch{$^8$Li$^+$}}

\newcommand{\rno}{\ch{$R$NiO$_{3}$}}
\newcommand{\lno}{\ch{LaNiO$_{3}$}}
\newcommand{\lao}{\ch{LaAlO$_{3}$}}

\DeclareSIUnit\unitcell{u.c.}
\DeclareSIUnit\gauss{G}

\newcommand\Tstrut{\rule{0pt}{2.9ex}}         
\newcommand\Bstrut{\rule[-1.2ex]{0pt}{0pt}}   
\newcommand\TBstrut{\Tstrut\Bstrut}           

\begin{document} 

\title{Evolution of the metallic state in LaNiO\texorpdfstring{$_3$}{3}/LaAlO\texorpdfstring{$_3$}{3} superlattices measured by \texorpdfstring{$^8$}{8}Li \texorpdfstring{$\beta$}{beta}-detected NMR}

\author{Victoria L. Karner}
\email[Email: ]{vkarner@chem.ubc.ca}
\affiliation{Department of Chemistry, University of British Columbia, Vancouver, BC V6T~1Z1, Canada}
\affiliation{Stewart Blusson Quantum Matter Institute, University of British Columbia, Vancouver, BC V6T~1Z4, Canada}

\author{Aris Chatzichristos}
\affiliation{Stewart Blusson Quantum Matter Institute, University of British Columbia, Vancouver, BC V6T~1Z4, Canada}
\affiliation{Department of Physics and Astronomy, University of British Columbia, Vancouver, BC V6T~1Z1, Canada}

\author{David L. Cortie}
\altaffiliation{Current address: Institute for Superconducting and Electronic Materials, Australian Institute for Innovative Materials, University of Wollongong, North Wollongong, NSW 2500, Australia}
\affiliation{Department of Chemistry, University of British Columbia, Vancouver, BC V6T~1Z1, Canada}
\affiliation{Stewart Blusson Quantum Matter Institute, University of British Columbia, Vancouver, BC V6T~1Z4, Canada}
\affiliation{Department of Physics and Astronomy, University of British Columbia, Vancouver, BC V6T~1Z1, Canada}

\author{Derek Fujimoto}
\affiliation{Stewart Blusson Quantum Matter Institute, University of British Columbia, Vancouver, BC V6T~1Z4, Canada}
\affiliation{Department of Physics and Astronomy, University of British Columbia, Vancouver, BC V6T~1Z1, Canada}

\author{Robert F. Kiefl}
\affiliation{Stewart Blusson Quantum Matter Institute, University of British Columbia, Vancouver, BC V6T~1Z4, Canada}
\affiliation{Department of Physics and Astronomy, University of British Columbia, Vancouver, BC V6T~1Z1, Canada}
\affiliation{TRIUMF, 4004 Wesbrook Mall, Vancouver, BC V6T~2A3, Canada}

\author{C. D. Philip Levy}
\affiliation{TRIUMF, 4004 Wesbrook Mall, Vancouver, BC V6T~2A3, Canada}

\author{Ruohong Li}
\affiliation{TRIUMF, 4004 Wesbrook Mall, Vancouver, BC V6T~2A3, Canada}

\author{Ryan M. L. McFadden}
\altaffiliation{Current address: TRIUMF, 4004 Wesbrook Mall, Vancouver, BC V6T~2A3, Canada}
\affiliation{Department of Chemistry, University of British Columbia, Vancouver, BC V6T~1Z1, Canada}
\affiliation{Stewart Blusson Quantum Matter Institute, University of British Columbia, Vancouver, BC V6T~1Z4, Canada}

\author{Gerald D. Morris}
\affiliation{TRIUMF, 4004 Wesbrook Mall, Vancouver, BC V6T~2A3, Canada}

\author{Matthew R. Pearson}
\affiliation{TRIUMF, 4004 Wesbrook Mall, Vancouver, BC V6T~2A3, Canada}

\author{Eva Benckiser}
\affiliation{Max Planck Institute for Solid State Research, 70569 Stuttgart, Germany}

\author{Alexander V. Boris}
\affiliation{Max Planck Institute for Solid State Research, 70569 Stuttgart, Germany}

\author{Georg Cristiani}
\affiliation{Max Planck Institute for Solid State Research, 70569 Stuttgart, Germany}

\author{Gennady Logvenov}
\affiliation{Max Planck Institute for Solid State Research, 70569 Stuttgart, Germany}

\author{Bernhard Keimer}
\affiliation{Max Planck Institute for Solid State Research, 70569 Stuttgart, Germany}

\author{W. Andrew MacFarlane}
\email[Email: ]{wam@chem.ubc.ca}
\affiliation{Department of Chemistry, University of British Columbia, Vancouver, BC V6T~1Z1, Canada}
\affiliation{Stewart Blusson Quantum Matter Institute, University of British Columbia, Vancouver, BC V6T~1Z4, Canada}
\affiliation{TRIUMF, 4004 Wesbrook Mall, Vancouver, BC V6T~2A3, Canada}

\date{\today}

\newcommand{\latin}[1]{\emph{#1}}

\begin{abstract}
Using ion-implanted \eli\ $\beta$-detected NMR, we study the evolution of the correlated metallic state of \lno\ in a series of \lno/\lao\ superlattices as a function of bilayer thickness. 
Spin-lattice relaxation measurements in an applied field of 6.55 T reveal two equal amplitude components: one with metallic ($T$-linear) $1/T_{1}$, and a second with a more complex $T$-dependence. 
The metallic character of the slow relaxing component is only weakly affected by the \lno\ thickness, while the fast component is much more sensitive, exhibiting the opposite temperature dependence (increasing towards low $T$) in the thinnest, most magnetic samples. 
The origin of this bipartite relaxation is discussed.
\end{abstract}

\maketitle

The perovskite rare-earth nickelates (\rno) are an important example of a metal-insulator transition (MIT) in a strongly correlated system\cite{Catalano2018}.
The transition can be tuned by the rare-earth cation \ch{$R$^{3+}} which modifies the \ch{Ni-O-Ni} angle\cite{Catalano2018}, and hence, the hopping integral and conduction bandwidth. 
\lno\ is the only \rno\ with a metallic ground state at all temperatures\cite{Catalano2018}. 
Advances in epitaxial film growth have enabled synthesis of high quality films and heterostructures of \lno\cite{Boris2011}.
In superlattices (SLs) with insulating interlayers of \lao, it was found that a MIT and N\'eel order could be induced when the thickness of the \lno\ layers was decreased to \num{2} unit cells (u.c.)\cite{Boris2011,Frano2013}.
This motivated the current experiment to understand how the correlated metallic state in \lno\ changes as it approaches this thickness-controlled MIT.  

NMR is an essential tool for studying strongly correlated electron materials, such as the high-$T_{c}$ cuprates\cite{Walstedt2008}; however, it is limited to bulk samples with native NMR nuclei. 
An alternative is $\beta$-detected NMR (\bnmr)\cite{2015-MacFarlane-SSNMR-68-1}, where a spin-polarized radioisotope is implanted into the sample and the resulting $\beta$-decay is used to detect the NMR
similar to muon spin rotation ($\mu$SR). 
\bnmr\ is a powerful tool for studying metals\cite{2015-MacFarlane-SSNMR-68-1}, including strongly correlated \ch{Sr2RuO4}\cite{Cortie2015} and \lno\cite{Karner2019}.

Here, \bnmr\ is used to study the evolution of the metallic state of \lno\ with decreasing thickness in a series of \lno/\lao\ SLs.  
Spin-lattice relaxation (SLR) measurements reveal two components: one with $T$-linear relaxation below \SI{200}{\kelvin}, and a second with a more complex $T$-dependence.
The slow component retains a metallic character similar to bulk \lno, independent of thickness.
In contrast, the fast relaxing component depends on thickness, deviates strongly from metallic, and exhibits a low $T$ upturn, consistent with the emergence of magnetic ordering.

{\it Samples:} Using pulsed laser deposition, the SLs were deposited on \ch{(LaAlO3)$_{0.3}$(SrAlTaO6)$_{0.7}$} (LSAT) substrates and annealed in an \ch{O2} rich atmosphere, similar to Ref.\ \onlinecite{Boris2011}.
The deposition begins with $n$ ($=$ \numlist[list-final-separator = {, }]{2;3;4;10}) \si{\unitcell} of \lno\ followed by an insulating interlayer of \SI{2}{\unitcell} of \lao.
The stacking sequence ($n$||$2$) was repeated \num{20} times for each SL (see \Cref{tab:sample-list}), terminating with \lao.
The crystallinity, interface sharpness, and layer thicknesses were verified using X-ray diffraction. 
Prior to the \bnmr\ experiments, the samples (\SI[product-units = power]{5x5x0.5}{\milli\meter}) were mounted on \ch{Al2O3} crystals with \ch{Ag} paint for compatibility with the spectrometer's cold-finger cryostat.
More details are given in the Supporting Information.

 \begin{table}[h!]
 	\centering
 	\caption{Summary of the Superlattice Samples}
 	\label{tab:sample-list}
 	\begin{tabular}{ r  l  l  l }
 		\hline
 		\hline
 		$n$ \TBstrut& Composition \TBstrut & Bilayer Thickness \TBstrut & Total Thickness \TBstrut \\ 
 		\hline 
 		\num{10}\Tstrut & \ch{[(LaNiO3)$_{10}$(LaAlO3)$_2$]$_{20}$} \Tstrut & \SI{4.6+-0.5}{\nano\meter} \Tstrut & \SI{92+-1}{\nano\meter} \Tstrut \\ [0.5ex]
 		\num{4} & \ch{[(LaNiO3)$_{4}$(LaAlO3)$_2$]$_{20}$} & \SI{2.6+-0.5}{nm} & \SI{52+-1}{\nano\meter} \\ [0.5ex]
 		\num{3} & \ch{[(LaNiO3)$_{3}$(LaAlO3)$_2$]$_{20}$} & \SI{2.0+-0.5}{\nano\meter} & \SI{40+-1}{\nano\meter} \\ [0.5ex]
 		\num{2}\Bstrut & \ch{[(LaNiO3)$_{2}$(LaAlO3)$_2$]$_{20}$}\Bstrut & \SI{1.5+-0.5}{\nano\meter} \Bstrut & \SI{30+-1}{\nano\meter} \Bstrut \\[0.5ex]
 		\hline
 		\hline
 	\end{tabular}
 \end{table}

{\it Methods:} To measure the \bnmr, a spin-polarized radioactive \elip\ ion beam was implanted into the sample. The measured $\beta$-decay asymmetry is proportional to the average longitudinal nuclear spin-polarization\cite{1983-Ackermann-TCP-31-291,2015-MacFarlane-SSNMR-68-1}, with a proportionality constant $A_0$ depending on the detector geometry and decay properties.
The asymmetry was obtained by combining the count rates in two opposing scintillation detectors placed along the polarization axis.
The experiments were performed at the ISAC
facility at TRIUMF in Vancouver, Canada\cite{2015-MacFarlane-SSNMR-68-1}.
The NMR isotope, \eli\ has nuclear spin $I =$ \num{2}, gyromagnetic ratio $\gamma/2\pi =$ \SI{6.3016}{\mega\hertz \per \tesla}, nuclear electric quadrupole moment $Q =$ \SI[retain-explicit-plus]{+32.6}{\milli\barn}, and radioactive lifetime $\tau =$ \SI{1.21}{\second}. The implantation energy was \SI{4.9}{\kilo\electronvolt}, corresponding to a mean depth of \SI{\sim 21}{\nano\meter} with a straggle of \SI{11}{\nano\meter}\cite{Ziegler2010}.

To measure the SLR, the asymmetry was monitored both during and after the \SI{4}{\second} beam pulse, during which it approaches a dynamic steady-state, while afterwards it relaxes to near zero. 
Unlike conventional NMR, the \eli\ is hyperpolarized in-flight by optical pumping, so no radio frequency (RF) field is required. 
The SLR rates were measured from 5 to 300~K in an applied field of $B_0=$ \SI{6.55}{\tesla} normal to the surface.

\begin{figure}[ht]
	\includegraphics[width = 0.9\columnwidth]{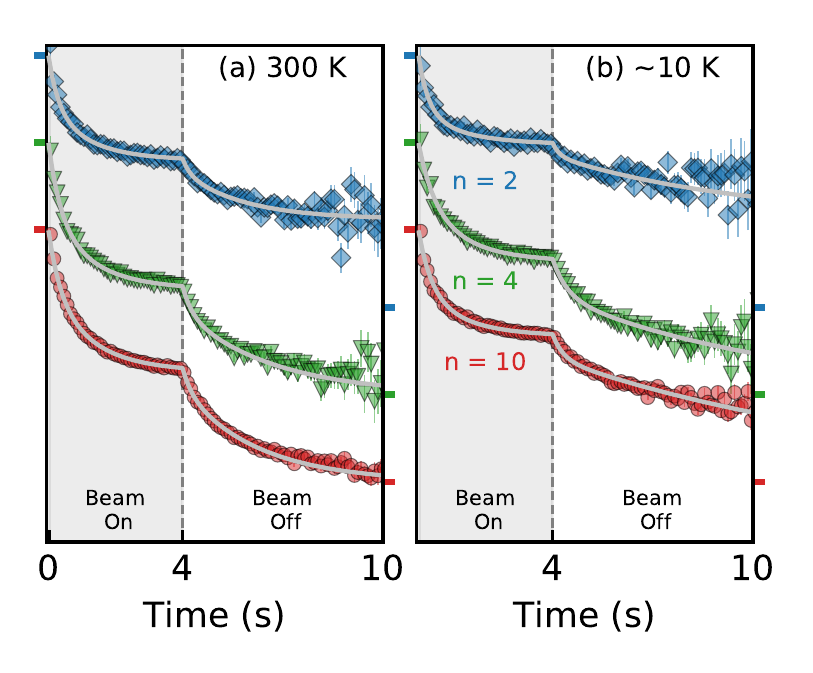}
	\caption{
		Examples of \elip\ SLR data in \ch{[(LaNiO3)$_n$(LaAlO3)$_2$]$_{20}$} with $B_{0} =$ \SI{6.55}{\tesla} at two temperatures for $n =$ \numlist[list-final-separator = {, and }]{2;4;10}.
		The lines are fits to \cref{eq:fit-fx-high-field} convoluted with the \SI{4}{\second} beam pulse indicated by the shaded area. 
		For clarity, the data has been offset vertically.  
		The coloured y-axis ticks on the left (right) represent the full (zero) asymmetry for each SL.
	}
	\label{fig:LNO-LAO-FitEx}
\end{figure}

{\it Results:} \Cref{fig:LNO-LAO-FitEx} compares the SLR data for different SLs.
The relaxation is easily measurable in all cases.
Generally, we expect it to be fastest in the (presumably most magnetic) $n=$ \num{2} SL, and slower for the more metallic samples (larger $n$). 
To quantify its evolution, a model relaxation function is required. 
The simplest choice that provides a good fit is a biexponential, to which we add a non-relaxing term
to account for a small fraction of \eli\ stopping in the substrate due to range straggling.
Specifically, at time $t$ after an \eli\ arriving at time $t'$, the polarization follows
\begin{equation} \label{eq:fit-fx-high-field}
R(t,t') = f_{\mathrm{SL}}[ (1-f_{f})e^{-\lambda_{s}(t-t')} + f_fe^{-\lambda_{f}(t-t')}] + (1-f_{\mathrm{SL}})
\end{equation}
where $\lambda_{i}$ ($i = s, f$) are the SLR rates. $f_{\mathrm{SL}}$ is the fraction of \eli\ in the SL, while $(1-f_{\mathrm{SL}})$ is in the substrate. Within the SL, $f_f$ is the fast relaxing fraction and $(1-f_f)$ is slow relaxing. The only $T$-dependent parameters in the fits were the two rates.
Biexponential relaxation appears intrinsic to \eli\ in \lno, and a similar analysis was necessary for a bulk single crystal\cite{Karner2019}. 
Interestingly, the fast relaxing fraction $f_{f} =$ \SI{50}{\%} independent of \lno\ thickness, and in agreement with the bulk\cite{Karner2019}. For further details on the analysis, see the Supporting Information.

\begin{figure*}[ht]
	\centering
	\includegraphics[width = 0.8\linewidth]{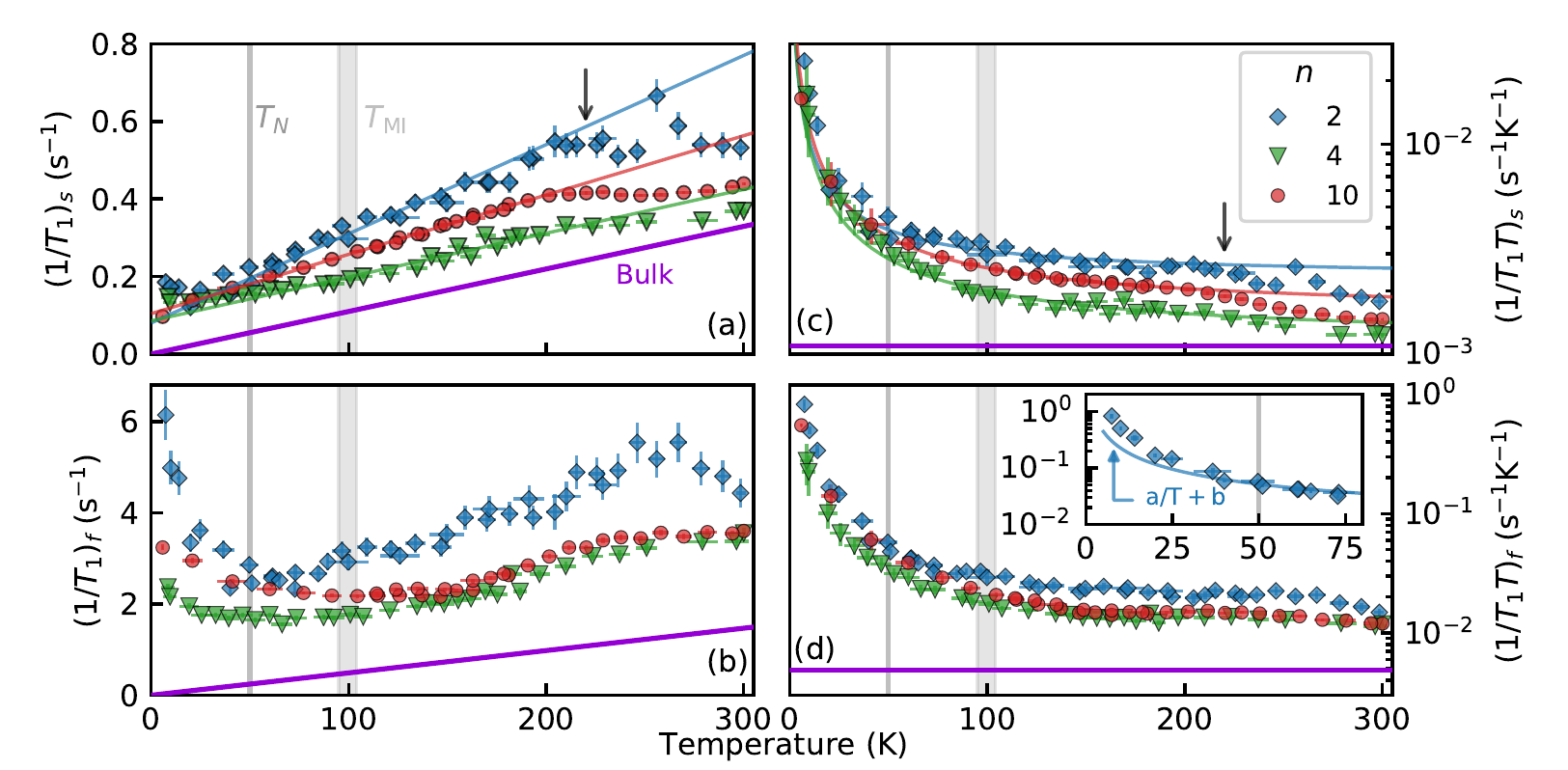}
	\caption{
		The slow (a),(c) and fast (b),(d) \eli\ SLR rates and $(T_{1}T)^{-1}$ as a function of temperature at $B_{0} =$ \SI{6.55}{\tesla} in \ch{[(LaNiO3)$_n$(LaAlO3)$_2$]$_{20}$} for $n =$ \numlist[list-final-separator = {, and }]{2;4;10}.
		Below \SI{\sim 200}{\kelvin}, $(1/T_{1})_{s}$ appears linear with a non-zero intercept (fits shown with solid lines).
		The vertical arrow at \SI{\sim 200}{\kelvin} indicates the sublinear deviation. 
		The shaded region and the vertical line illustrate the reported $T_{\mathrm{MI}}$ and $T_{N}$ in an $n =$ \num{2} SL on \ch{SrTiO3}\cite{Boris2011}. 
		For reference, the bulk Korringa slope is shown as the solid purple line\cite{Karner2019}.
		The inset of (d) illustrates the low-$T$ upturn for $n =$ \num{2}, which spans more than an order of magnitude. 
		The data for the $n =$ \num{3} SL has been omitted to avoid clutter; however, the trend is very similar to $n = 2$.
		}
	\label{fig:LNO-LAO-SLR-Tdep}
\end{figure*}

\begin{figure}[ht]
	\includegraphics[width = 0.75\columnwidth]{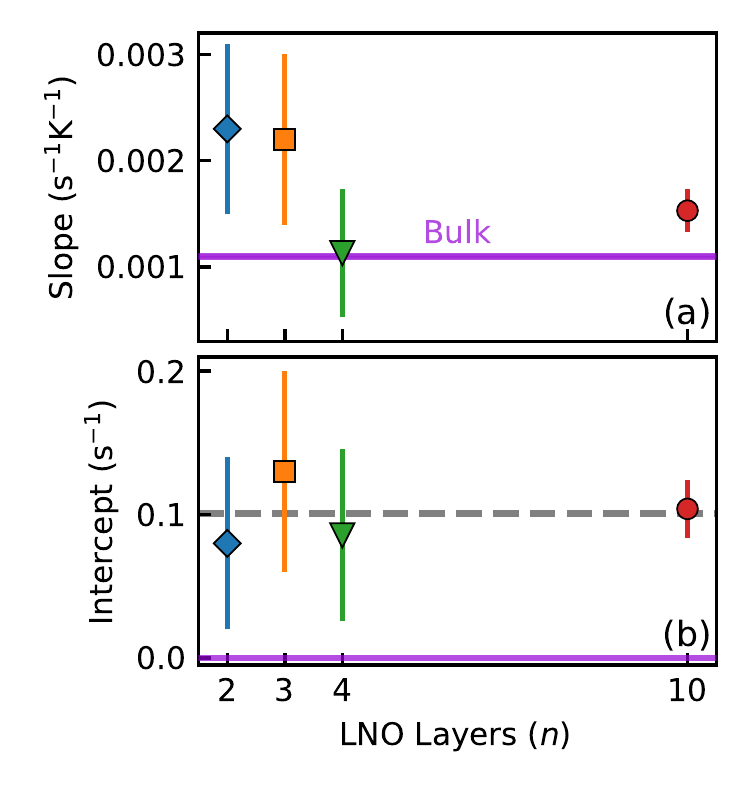}
	\caption{
	(a) Slopes from the linear region of the slow component as a function of the \lno\ thickness $n$. The bulk slope is shown as the solid purple line. The $n =$ \numlist{4;10} values are scattered around the bulk, while the thinner SLs ($n =$ \numlist{2;3}) are significantly enhanced.
	(b) The $T\rightarrow 0$ intercepts appear independent of $n$, with average value shown by the dashed grey line.
	}
	\label{fig:LNO-LAO-slopes}
\end{figure}

The resulting rates $(1/T_{1})_{i}\equiv \lambda_i$ are shown as a function of temperature in \Cref{fig:LNO-LAO-SLR-Tdep}. 
The slow rate in panel (a) remains linear (metallic) for all the SLs, with a slope similar to bulk \lno\cite{Karner2019}.
Closer inspection reveals that: \num{1}) above \SI{\sim 200}{\kelvin} (marked by the vertical arrow), $1/T_1$ becomes sublinear; \num{2}) the SLs are all faster relaxing than the bulk, and this enhancement is mostly due to a finite $n$-independent intercept as $T\rightarrow 0$; and \num{3}) the slope changes systematically with $n$, but not monotonically, as the shallowest slope is at $n =$ \num{4}.
Below \SI{200}{\kelvin}, we fit $(1/T_{1})_{s}$ to a line to obtain the values shown as a function of $n$ in \Cref{fig:LNO-LAO-slopes}.

In contrast, the fast rate in \Cref{fig:LNO-LAO-SLR-Tdep}(b) is much more strongly enhanced over the bulk (by a factor of \numrange[range-phrase = --]{3}{5} at \SI{200}{\kelvin}).
Similarly, there is a linear region (above \SI{\sim 100}{\kelvin}), with a slope not far from the bulk, but this linearity does not persist to low $T$. 
Instead, it shows an upturn below \SI{50}{\kelvin}, which is most dramatic at $n =$ \num{2}.
At the MIT, we expect to lose the metallic Korringa relaxation, but as in other \ch{$R$NiO3}, the insulating state contains Ni local moments. The increasing SLR at low $T$ is consistent with relaxation from fluctuations of the \ch{Ni} spins which slow and eventually freeze. 

Remarkably, we find no obvious signal from the \lao. 
Like LSAT\cite{LSAT-2017}, the relaxation in bulk \lao\ is extremely slow\cite{LAO-2017}. 
However, the \lao\ is very thin, and any \eli\ at the interface will retain some coupling to the adjacent \lno, making the distinctly \lao\ volume fraction rather small.
The \elip\ site energy in the \lao\ layer may also be higher than \lno\ due to the higher valent \ch{Al^{3+}}, resulting in a bias against stopping in this layer. 
In the thin SLs, the \lao\ fraction may be included in the substrate term, while in the thicker SLs it may make a small contribution to the slow component.

{\it Discussion:} Having outlined the results, we now discuss their implications.
Like the implanted muon in $\mu$SR, \elip\ is typically located at a high symmetry interstitial site in the host lattice. 
In perovskite oxides, this site (denoted $P$) is midway between adjacent $A$-site ions at the center of a square \ch{$B$O} plaquette\cite{MacFarlane2003,LSAT-2017,Karner2019}.
Analogous to the implanted muon and native NMR nuclei, \eli\ is coupled via the hyperfine interaction to the electronic system of the host. 
However, it is much longer lived than the muon, making it sensitive to phenomena on longer timescales such as Korringa relaxation.

With only one site, biexponential relaxation is unexpected.
In bulk \lno, where both rates are $T$-linear, we attributed it to two subtly different $P$-sites in an unconventional distorted perovskite structure\cite{Karner2019}. 
This was reasonable, since the rates differed only in magnitude, not temperature dependence.
However, it is incompatible with the present data where the two components exhibit distinct $T$ dependences.  With no evidence for other sites, we instead propose that the electronic state of \lno\ (even in a metallic single crystal) is \emph{intrinsically inhomogeneous}, i.e., microscopically separated into two equally abundant phases with distinct electronic properties.

In the Kondo lattice compound \ch{YbRh2Si2}, \ch{^{29}Si} NMR revealed a similar phase separation into distinct metallic states in the vicinity of the field induced quantum critical point\cite{Kambe2014}. Interestingly, the phase fraction (and not the relaxation rate) evolved with magnetic field and temperature, but converged toward $f_f =$ \num{0.5} ($R =$ \num{1} in their notation) at the lowest temperatures. 
\lno\ has been suggested to have a Kondo lattice character with a much larger energy scale\cite{Anisimov1999}. 
The phase separation revealed in our data, persists at least up to \SI{300}{\kelvin}, even in the single crystal; so, if it is analogous, we cannot expect its properties to be as purely electronic as the low-$T$ \ch{YbRh2Si2}, and lattice dynamics should play some role.

Phase separation between metallic and insulating phases is known in other \rno\cite{Mattoni2016,Preziosi2018}, but only near the MIT.
There, the insulating volume fraction varies from \numrange{0}{1} through a narrow coexistence region around the transition. 
In contrast, our volume fraction is constant over the whole temperature range, including in the metallic limit of the single crystal\cite{Karner2019}. 
If this phenomenon is related, then the metal-insulator phase separation must reflect some underlying inhomogeneity present in the high temperature metallic phase. 
This is likely structural in origin, as suggested by its correlation with step-edges, and the stability of its microstructure through temperature cycling\cite{Mattoni2016}.
Candidate sources of structural heterogeneity are competing distorted perovskite structures
(octahedral rotation pattern) or bond disproportionation (octahedral size)\cite{Alonso1999PRL,Catalano2018}. 
We note the precise structure of \lno\ remains elusive. Although it is considered rhombohedral, recent neutron scattering results indicate a lower local symmetry\cite{Li2015,Shamblin2018}.
It has also been suggested to contain nanoscopic monoclinic insulating pockets\cite{Li2015}, which is inconsistent with the two distinct metallic phases we find in bulk\cite{Karner2019}.  

In summary, the biexponential relaxation strongly suggests \emph{microscopic electronic phase separation} in \lno\ that persists up to at least \SI{300}{\kelvin}.
The phase separation is static on the second time scale and is robust across multiple \lno\ samples, including single crystal\cite{Karner2019}, thin film\cite{Karner2019}, and SLs.
Confirming it may require other local probes, such as conventional NMR, or scanning probe methods that could directly image its spatial distribution\cite{Preziosi2018}.

In similar SLs, low energy $\mu$SR (in zero and low fields) finds a broad distribution of static fields (150 G width) below \SI{50}{\kelvin} at $n =$ \num{2} which pervade the entire sample volume\cite{Boris2011} , seemingly inconsistent with our results.
However, $1/T_1$ does not probe the static fields directly, instead being determined by field fluctuations at the Larmor frequency (\SI{41.27}{\mega\hertz}).
On the other hand, static fields could propagate out from an incomplete magnetic volume fraction into the nonmagnetic phase, provided the two were intimately mixed at the nanoscale. 
This may explain why the scale of the internal fields is more than $10\times$ smaller than bulk \rno\cite{Frandsen2016}. 
Alternatively, in analogy with \ch{YbRh2Si2}, we may have stabilized a distinct phase-separated ground state with the large $B_0$ field.

Now we turn to what the evolution of $1/T_1$ can tell us about the MIT at small $n$.
In a conventional metal, the Korringa $T$-linear dependence is a consequence of Fermi statistics and a density of states (DOS) that is smooth and featureless on the scale of $kT$.
The Korringa slope is then proportional to the square of the DOS at the Fermi level $\rho(E_F)$.
Considering \Cref{fig:LNO-LAO-slopes}(a), there is a 2-fold increase in the slow component's slope for $n <$ \num{4}, consistent with an effective narrowing of the conduction band on approaching the 2D limit. 
However, for all $n$, we find a finite intercept, inconsistent with simple metallic behaviour.
More generally, for a correlated metal susceptible to magnetic order, the Moriya expression\cite{Moriya1963} relates $1/T_{1}$ to the imaginary part of the generalized susceptibility $\chi(\boldsymbol{q},\omega)$, 
\begin{equation}
\frac{1}{T_{1}T} = \frac{4k_{B}}{\hbar} \sum_{\boldsymbol{q}}^{} \frac{\lvert A(\boldsymbol{q}) \rvert ^{2}}{\left(\gamma_{e}\hbar\right)^{2}} \frac{1}{\hbar \omega_{0}} \chi''\left(\boldsymbol{q},\omega_{0}\right),
\label{Moriya}
\end{equation}
where $\gamma_{e}$ is the electron gyromagnetic ratio, $\omega_{0}$ is the NMR frequency, and $A\left(\boldsymbol{q}\right)$ is the hyperfine form factor which acts as a filtering function on magnetic fluctuations at wavevector $\boldsymbol{q}$. 
For a conventional metal, the summation in \Cref{Moriya} is independent of $T$ for $kT \ll E_F$.
A finite intercept implies that, in addition to a constant part, the sum contains a term proportional to $1/T$.
This is seen clearly in the corresponding plot of $1/(T_1T)$ (on a log scale) in \Cref{fig:LNO-LAO-SLR-Tdep}(c).
Such a Curie-like term suggests a population of dilute uncoupled local moments, as has been considered near the MIT in doped semiconductors\cite{Gan1986}.
Interestingly, this term appears independent of $n$, so it is probably not caused by \ch{Ni} moments at the \lao\ interface.

Aside from the intercept, the slow relaxing phase exhibits a metallic character that is not very strongly modified from the bulk, even for the smallest $n$.
In contrast, the fast relaxing phase exhibits a much more dramatic sensitivity to thickness.
In the bulk, the factor of \num{4} between the Korringa slopes in the fast and slow phases\cite{Karner2019} would imply a factor of \num{2} in $\rho(E_F)$ in an uncorrelated metal.
In the SLs, compared to the slow component, $(1/T_1)_f$ shows a $10\times$ larger $T$-independent term (vertical shift) and an upturn below \SI{\sim 50}{\kelvin} with $(1/T_1)_f$ doubling between \SI{50}{\kelvin} and \SI{4}{\kelvin} at $n=2$ (inset of \Cref{fig:LNO-LAO-SLR-Tdep}(d)).
Similar $n =$ \num{2} SLs have a zero field N\'eel transition at $T_{N}$ \SI{\sim 50}{\kelvin}\cite{Boris2011,Frano2013} near the upturn and an ordering wavevector $\boldsymbol{q}_{\mathrm{AF}} = (1/4,1/4,1/4)$\cite{Frano2013}. The $P$ site hyperfine form factor $A(\boldsymbol{q}_{\mathrm{AF}})\neq 0$, so \eli\ senses fluctuations at $\boldsymbol{q}_{\mathrm{AF}}$.
However, even the thicker SLs with $n =$ \numlist{4;10} (that are thought to remain nonmagnetic and metallic) show a similar, albeit muted feature.
The upturn implies a term in \Cref{Moriya} increasing more steeply than $1/T$ at low $T$.
This is likely related to magnetic ordering, but probably modified by the strong applied field (cf.\ the transition in \ch{Mn}-doped \ch{Bi2Te3}\cite{McFadden2020}).
Note the vertical scale in \Cref{fig:LNO-LAO-SLR-Tdep}(d) spans an order of magnitude more range than the slow component in \Cref{fig:LNO-LAO-SLR-Tdep}(c).

Finally, we return to the sublinearity in $(1/T_{1})_{s}$ above \SI{200}{\kelvin} (vertical arrow in \Cref{fig:LNO-LAO-SLR-Tdep}). The deviation is subtle, sample dependent, and does not vary monotonically with $n$, as it is smallest at $n=4$.
A similar deviation was observed in bulk \lno, although there it had the opposite sense for the slow and fast components.
This feature is probably related to small changes in the \ch{Ni-O-Ni} angle around \SI{200}{\kelvin}\cite{Li2015} which also coincides with a decrease in the carrier density from RF conductivity\cite{Shamblin2018}.
In addition, the magnetic susceptibility
crosses over from a low temperature Pauli-like regime to a high temperature Curie-Weiss dependence\cite{Zhang2017} at about \SI{200}{\kelvin}. It is reasonable to expect such a small structural change would differ between a crystal, SL, and film, and it may even depend on the precise thermal history.

{\it Conclusion:}
\bnmr\ spin-lattice relaxation measurements of implanted \eli\ in a series of \lno/\lao\ SLs reveal biexponential relaxation, consistent with single crystal \lno\cite{Karner2019}; but, in contrast, the two components exhibit very different temperature dependences.
Thus, we propose \lno\ is electronically phase separated.
The slow relaxing phase is metallic, not strongly modified from the bulk, and persists to low $T$, even for the thinnest ($n =$ \num{2}) SL.
As $n \rightarrow$ \num{2}, the slope and the DOS are enhanced consistent with band narrowing towards the 2D limit. 
In contrast, the fast relaxing phase is much more sensitive to $n$ and appears magnetic at low $T$. 

{\it Acknowledgments:} We thank R. Abasalti, D. J. Arseneau, S. Daviel, B. Hitti, and D. Vyas for technical assistance.
We thank A. Frano, S. Johnston, Q. Si, and M. Vojta for useful discussions. 
This work was supported by NSERC Canada.
Work at the MPI-FKF (sample synthesis and characterization) was supported by the Deutsche Forschungsgemenischaft (DFG, German Research Foundation) - Projektnummer 107745047 - TRR 80.


%

\end{document}